# The intelligent prediction and assessment of financial information risk in the cloud computing model

Yufu Wang[1]*, Mingwei Zhu[1.2], Jiaqiang Yuan[2], Guanghui Wang[3], Hong Zhou[4]

1*Computer Science & Engineering, Santa Clara University, Santa Clara, CA, USA
Corresponding author: ywang42@scu.edu
1.2 Computer Information System, Colorado state university, Fort Collins, CO, USA
2 Information Studies, Trine University, phoenix, AZ
3 Computer Science, Independent Contributor, Shanghai, CN
4 Computer Technology, Peking University, Beijing, CN

**Abstract.** cloud computing (cloud computing) is a kind of distributed computing, referring to the network "cloud" will be a huge data calculation and processing program into countless small programs, and then, through the system composed of multiple servers to process and analyze these small programs to get the results and return to the user. This report explores the intersection of cloud computing and financial information processing, identifying risks and challenges faced by financial institutions in adopting cloud technology. It discusses the need for intelligent solutions to enhance data processing efficiency and accuracy while addressing security and privacy concerns. Drawing on regulatory frameworks, the report proposes policy recommendations to mitigate concentration risks associated with cloud computing in the financial industry. By combining intelligent forecasting and evaluation technologies with cloud computing models, the study aims to provide effective solutions for financial data processing and management, facilitating the industry's transition towards digital transformation.

**Keywords:** Cloud Computing, Financial Information Processing, Intelligent Technologies, Data Security, Regulatory Frameworks

## 1. Introduction

Applying the cloud computing model to financial information processing has become a major trend in today's financial industry. With the explosive growth of incoming data, financial institutions are facing unprecedented data challenges. Traditional data processing relies on manual review, which leads to low business efficiency and easy to miss business opportunities. There is an urgent need for financial institutions to adopt intelligent technologies to address this challenge and improve the efficiency and accuracy of data processing. [1]According to the National Institute of Standards and Technology (NIST), cloud computing is a pay-per-use model, through the cloud computing, Users can obtain network, server, storage, application and other resources from the configurable computing resource sharing pool whenever and wherever needed. These resources can be provisioned and released quickly, minimizing administrative effort and service provider involvement.

Cloud services have become an important part of the global financial industry's information technology toolbox. With the increasing use of cloud services by financial institutions, financial regulators have raised concerns about the potential concentration risks posed by cloud services. Against the background of the application of cloud technology in the financial industry, this report highlights the potential risks of financial institutions' use of third-party technology service providers, draws on the regulatory frameworks of different jurisdictions, and proposes policy recommendations to enhance information gathering and sharing, strengthen cross-border coordination and solutions, and ensure that regulatory tools are fit for purpose. To mitigate the potential concentration risks posed by cloud computing applications in the financial industry.

As a result, financial information in the cloud computing environment faces many risks and challenges. Traditional data processing methods have found it difficult to meet the growing data needs and processing requirements of the financial industry. Especially in the aspect of financial data entry and analysis, the traditional OCR[2] model has many limitations, which cannot meet the requirements of high efficiency and high accuracy of data processing by financial institutions. In addition, data security, privacy and other issues in the cloud computing environment also bring new challenges to financial information processing.

This study aims to solve the risks and challenges of financial information in the cloud computing environment by using intelligent forecasting and evaluation technology combined with the cloud computing model. The introduction of new digital applications and cloud services, as well as the use of artificial intelligence and other related technologies, will improve the efficiency and accuracy of financial institutions' data processing and accelerate the digital transformation of financial enterprises. Through such research, more effective data processing and management solutions can be provided to the financial industry, and the entire industry can be promoted towards intelligent and digital transformation.

## 2. Related work

*2.1. Cloud computing model*

Cloud computing is a model that makes it easy to access a common set of configurable computing resources (such as networks, servers, storage devices, applications, and services) over a network. These resources can be made available and distributed quickly, with minimal administration costs and service provider intervention."More than a decade ago, cloud computing was considered an insurmountable technology by China's IT industry, but with the success of Alibaba Cloud's research and development, domestic giants' cloud computing has blossomed. Now cloud computing is everywhere, whether it is government, enterprise or consumer, we basically use different clouds every day. There are four main types of cloud computing: private cloud, public cloud, hybrid cloud and multi-cloud. At the same time, there are three main cloud computing service models: infrastructure as a service (IaaS), platform as a service (PaaS) and software as a service (SaaS), these three are the current mainstream cloud computing service models, the future, with the development of technology, cloud computing will develop into the fourth model: S2S (service to service). This is the International Institute of Electrical and Electronics Engineers [3](IEEE Fellow), the American Computer Society outstanding scientist Mr Zhang Liangjie put forward.

In terms of where computing takes place, cloud computing moves the running of software from the usual personal computer (or desktop computer) to the cloud, which is a server or cluster of servers located in some "mysterious" geographical location. These servers or clusters can be located locally, off-site or even far away. This seems to be the client/server model, but cloud computing is not the traditional client/server model, but a huge improvement on that model. From the form of resource supply, cloud computing is a service computing, that is, all IT resources, including hardware, software, architecture are sold and charged as a service. For cloud computing, there are three main services provided: Infrastructure-as-a-service (IaaS), which provides hardware resources similar to the traditional model of CPU, storage, and I/O; Platform as a Service (PaaS), which provides an environment for software to run, similar to the operating system and programming framework under the traditional programming model; Software as a Service (SaaS) provides application software functionality, similar to the traditional model of application software. In the cloud computing model, users no longer buy or buy some hardware, system software or application software to become the owner of these resources, but purchase the use time of resources, and consume according to the billing model of the use time.

It can be seen that cloud computing takes all resources as services and consumes them in a pay-as-you-go manner, which is the characteristic of the host era. [4]In the host era, all users are charged based on the CPU time and storage capacity consumed when they connect to the host through the display terminal and network cable. The difference is that in host mode, the computation takes place on a single host; In cloud computing, computing takes place in a cluster of servers or in a data center.

*2.2. Features of cloud computing environment*

1) High degree of resource integration. Compared with the local computing environment, the resource utilization level obtained by a single user in the cloud computing platform may not be superior due to the network speed and other reasons, but the utilization rate of some idle resources has been greatly improved, thus maximizing the utilization rate of limited overall resources and greatly improving the resource utilization rate of the whole society.

2) Strong impact resistance. The cloud platform adopts the distributed data storage mode, which not only provides the basis for data recovery, but also makes various network attacks become confused, and plays an important role in improving the security and anti-impact capability of the system.

3) High scalability. The cloud platform adopts a modular design. At present, the mainstream cloud computing platform integrates hardware and software devices and middleware software with different functions at each layer according to SP architecture. A large number of middleware software and devices provide a common interface to the platform, allowing users to add extended devices for this layer]

4) Low use cost. Due to the adoption of distributed data storage, cloud computing mode greatly saves hardware equipment acquisition costs, so that users can use the remaining idle funds to order according to their own planning and needs, and improve the utilization rate of funds.

*2.3. Traditional financial information security*

The development of the financial information security industry can be traced back to the 1980s, when with the popularization of computer technology and the electronization of financial services, financial institutions began to realize the importance of information security. In the early 1990s, with the rise of the Internet, the issue of financial information security became more prominent, and financial institutions began to take various measures to protect the security of customer information and transaction data. [5]At the beginning of the 21st century, with the rapid development of e-commerce and the rise of mobile payment, the financial information security industry has entered a stage of rapid development, and a number of professional financial information security companies and technical service providers have emerged. In recent years, with the application of new technologies such as artificial intelligence and big data, the financial information security industry is facing new challenges and opportunities, and it needs continuous innovation and development to cope with ever-changing threats.

The upstream of the financial information security industry is mainly basic hardware and software, of which the basic hardware includes chips, memory and other devices, and the basic software includes operating systems, databases and middleware. These infrastructures provide necessary technical support for financial information security. [6]At the application level, financial information security applies information security technologies such as cryptography, key management, identity authentication, access control, application security protocol and transaction processing into financial information system security engineering according to the actual application needs of financial system. In addition, the downstream application field is very wide, including not only traditional financial institutions such as banking, securities, insurance, etc., but also emerging financial technology companies.

As of June 2023, the number of Internet users in China has reached 1.079 billion, and the penetration rate is as high as 76.4%. At the same time, the scale of online payment users has continued to grow, reaching 943 million in June, an increase of 31.76 million compared with December 2022, accounting for 87.5% of all Internet users. With the widespread adoption of Internet use and online payments, digital identity information, as a core component of the digital finance industry, is facing unprecedented security risk challenges. These data contain a large amount of personal sensitive information, such as transaction records and user personal information. At the same time, the generation and accumulation of financial data is accelerating due to the growing demand for digital financial services. However, in the current environment, financial data security laws and regulations are not yet complete, resulting in frequent data breaches, tampering and extortion. These problems have spawned a huge illegal data trafficking market, which poses a serious threat to users' privacy and personal information security.

With the transformation of payment habits, China's Internet finance industry has experienced significant growth, which has also led to a surge in the number of Internet finance websites. These sites play a key role in the financial sector, as well as being an important segment of cybersecurity. According to the data released by the National Internet Finance Risk Analysis Technology Platform, as of October 2021, the number of Internet finance websites registered in China has totaled 77,932. It is worth noting that out of this huge number, more than a fifth (21,371) of the sites were found to have anomalies. In addition, the data shows that nearly three-quarters (76.08%) of all Internet finance platforms have ceased operations. This reflects the strict level of cybersecurity management of Internet financial platforms in China, as well as the importance regulators attach to safeguarding users' interests and financial stability.

*2.4. The application of cloud computing in the financial field*

1. Banking. Cloud computing is mainly used in IT operations management and open underlying platforms. The application of cloud computing technology to build an open cloud platform can build a comprehensive financial service ecosystem with the help of API, providing "financial + non-financial" services such as life payment, information inquiry, online shopping, and the combination of financial services and life scenarios enhances the value of financial accounts.
Objective: (1) To enhance data security - to promote the development of retail business, online service operation mode and customer demand personalized service; Enhance the storage capacity and reliability of bank data; (3) Reduce bank costs and improve bank operation efficiency.
2. Securities fund field. Cloud computing is mainly used in client market query and peak trading volume allocation. Through the business system as a whole in the cloud, under the deployment mode of database sub-database and sub-table, the parallel computing equivalent to thousands of clearing systems and real-time trading systems can be realized[7].
Take Shenyin Wanguo Securities Company as an example, the company completed the construction of enterprise cloud computing platform project in 2009 and brought new business and good economic benefits. The project has the following five features: (1) The virtualization technology is used to build a shared data center, which realizes the on-demand allocation of resources and the reliable processing of massive data; (2) Build a cloud computing trusted network platform based on the principle of multi-point redundancy and effective isolation; (3) Provide standardized business platform cloud services for the securities industry; (4) Architecture of high-performance application infrastructure platform cloud services; ⑤ Realized the deployment and operation of a variety of online application systems, and formed an enterprise cloud computing platform operation and maintenance management system characterized by unification, standardization and automation.

3. Insurance. Cloud computing is mainly used in personalized pricing and product online sales. Customized cloud software can quickly analyze real-time customer data, provide personalized pricing, and provide specialized insurance services to targeted customers through social media.

*2.5. Features of cloud computing environment*

1) High degree of resource integration. Compared with the local computing environment, the resource utilization level obtained by a single user in the cloud computing platform may not be superior due to the network speed and other reasons, but the utilization rate of some idle resources has been greatly improved, thus maximizing the utilization rate of limited overall resources and greatly improving the resource utilization rate of the whole society.

2) Strong impact resistance. [7]The cloud platform adopts the distributed data storage mode, which not only provides the basis for data recovery, but also makes various network attacks become confused, and plays an important role in improving the security and anti-impact capability of the system.

3) High scalability. The cloud platform adopts a modular design. At present, the mainstream cloud computing platform integrates hardware and software devices and middleware software with different functions at each layer according to SP architecture. A large number of middleware software and devices provide a common interface to the platform, allowing users to add extended devices for this layer]

4) Low use cost. Due to the adoption of distributed data storage, cloud computing mode greatly saves hardware equipment acquisition costs, so that users can use the remaining idle funds to order according to their own planning and needs, and improve the utilization rate of funds.

## 3. Methodology

*3.1. Experimental background*

By analyzing the online migration of financial business and the extensive penetration of Internet business model, this paper finds that business transaction scale and business peak per second show geometric growth. In this context, cloud native is gradually recognized and favored by business teams as an important tool for enterprise cloud. However, with the increasing prominence of financial information security risks, financial institutions urgently need to combine cloud-native technology with security measures to ensure the safety and reliability of financial information（figure 1）.

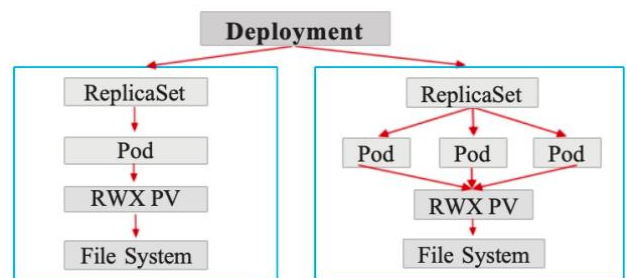

**Figure 1.** Deployment application implementation information storage architecture diagram

Cloud native not only can achieve cost reduction and efficiency, improve system resource utilization, but also has many advantages such as simple use, high availability, and strong expansion. It also has intelligent load, elastic scaling, log monitoring, [8]DevOps and other functions, can achieve multi-application on-demand selection and intelligent release, so as to cope with sudden business traffic, quickly achieve expansion and contraction capacity, greatly improve development efficiency, reduce operation and maintenance costs. However, with the continuous evolution of financial information security threats, financial institutions must strengthen the security control of cloud-native environments, including data encryption, access control, identity authentication and other security measures to ensure that financial data is not leaked or tampered with.

Container cloud combines service governance capabilities to achieve full lifecycle management of continuous integration and delivery of applications (figure2).

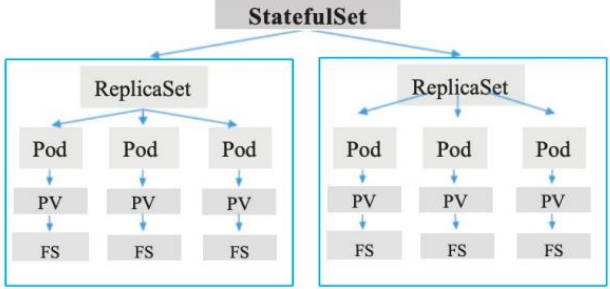

**Figure 2.** Statefulset Implements the information storage architecture diagram

In this process, financial institutions need to pay attention to information security governance, establish a sound security management system, including the formulation of security policies, strengthening security training, implementation of security audits and other measures to ensure that financial information in the cloud native environment is safe and controllable[9]. With the continuous verification of cloud native value in practice and the recognition of securities industry enterprises, financial institutions will also increase their investment in cloud native security, and security protection will be included in the important agenda of enterprise construction.

*3.2. Introduction to cloud native storage*

In the container cloud environment, the demand for intelligent prediction of financial information security is also very important. With the online migration of financial services and the widespread penetration of Internet business models, financial institutions are faced with increasingly complex information security challenges. Therefore, how to effectively use the storage method provided by the container cloud environment, combined with intelligent prediction technology, has become one of the important solutions for financial information security.

Because financial institutions can utilize a variety of storage methods to protect important financial data. For example, through the internal storage methods such as EmptyDir, HostPath[10], and Local PV, financial institutions can store and share temporary data. At the same time, combined with external CSI, NAS, and SAN storage methods, they can achieve secure storage and management of persistent data. On this basis, financial institutions can use intelligent prediction technology to make more accurate and effective prediction and evaluation of financial information security.

Therefore, intelligent prediction technology can use machine learning, data mining and other methods to analyze and mine financial data, so as to find potential security threats and risks. By continuously optimizing and improving intelligent prediction models, financial institutions can improve their ability to identify and respond to security threats and ensure the security and stability of financial information. Therefore, the combination of storage in the container cloud environment and intelligent prediction technology can provide financial institutions with a more comprehensive and effective information security solution.

*3.3. Container cloud nonpersistent storage requirements*

In the cloud native container cloud environment, the demand for intelligent prediction of financial information security is also critical. In this environment, there are usually some non-persistent storage of sensitive services, these services need to use the container application's image warehouse, running container instances, and generated logs and other data. Scenarios for storage needs of financial institutions in the container cloud include, but are not limited to:

**Table 1.** Cloud-Native Persistent Storage Requirements

| Requirement | Description |
| --- | --- |

| Requirement | Description |
|---|---|
| Persistent Demand | Essential for production system business logs and foundational for business analysis, requires permanent storage. |
| Pod Drift | After Pod migration, state data migration is achieved by mounting the same storage on another node. |
| Shared Storage | Distributed sharing requirements for business files, images, and other data. |
| Scalability | Flexible scalability requirements for storage, as well as flexible expansion and migration capabilities for container nodes. |
| Performance | Directly related to the concurrency support capability of business systems, especially for application files such as images and documents. |
| Security & High Availability | Requires intranet storage to ensure dual-center architecture. |

Container instance creation and running: In a container cloud environment, the creation and running of each container instance requires a certain amount of storage space. The size of these storage Spaces depends on the number of containers, the underlying environment in which the containers are deployed, and the specific requirements of the container application. The storage space of the overall container working node needs to be large enough to cover the total space requirements of all containers that may run on that node. For the storage space of running containers, high performance is required to support fast dynamic expansion of containers and fast read and write of files.

Temporary file storage: [11]During the running of the container application, some temporary files are generated, which do not require persistent storage. These temporary files require a certain amount of storage space to support the normal operation of the container application.

To meet the above storage requirements, financial institutions can select a proper non-persistent storage model based on specific service scenarios and performance requirements. Generally speaking, local storage and block storage can be used for scenarios with high IO and low latency. Local storage can be achieved by building hostpaths or using centralized block storage for higher IOPS and bandwidth. For local hard disks, you are advised to configure RAID to improve the reliability of the overall service system and reduce the impact of single point of failure.

**Table 2.** Technologies Distribution in Various Categories

| Category | Technologies | Percentage |
|---|---|---|
| Database | Cassandra, MariaDB, MongoDB, MySQL, Neo4j, PostgreSQL, Redis, etcd | 64.38% |
| Content Management | Drupal, CKAN, MediaWIKI, DNN, Joomla | 10.96% |
| Continuous Integration | Jenkins, GitLab, Maven, Puppet | 42.47% |
| Big Data | Hadoop, Hypertable, Mesos, Presto, Solr, Spark, Storm | 12.33% |
| Analysis & Search | Grafana, ElasticSearch, Prometheus, Kibana, Logstash | 10.96% |
| Web Services | NGINX, WordPress, Apache HTTP Server, Tomcat, httpd | 45.21% |
| Infrastructure | RabbitMQ, Memcached, Kafka, ZooKeeper, Node.js, NATS, WildFly | 9.59% |
| Development Tools | LAMP | 16.44% |
| AI | MxNet, PyTorch, TensorFlow Notebook, TensorFlow ResNet | 5.48% |
| Others | | 8.22% |

In the cloud native container cloud environment, financial information security intelligent prediction needs to fully consider the selection and performance requirements of storage schemes to ensure the safe storage of financial data and the effective realization of intelligent prediction. [12]By properly configuring and optimizing the storage system, financial institutions can improve their ability to identify and respond to security threats, thus ensuring the security and stability of financial information.

*3.4. Cloud native information storage*

When using cloud-native platforms, users often need to share the use of Kubernetes clusters (multi-tenancy) to simplify operations and reduce costs while meeting the needs of multiple teams and customers. While Kubernetes does not directly provide multi-tenancy itself, it does provide a number of features that can be used to support the implementation of multi-tenancy. In the tenant granularity, you can perform resource isolation and management, user authentication, and feature management. Cloud-native storage should also be combined with tenant rights mechanisms. Because cloud native is an application-centered software development method, it provides practitioners with an efficient, scalable and replicable way to maximize the ability of the cloud and play the value of the cloud. Securities companies also need to continuously improve their software research and development capabilities. We can build an [13]IT delivery model that is as agile and flexible as an Internet company, with rapid trial and error and rapid innovation.

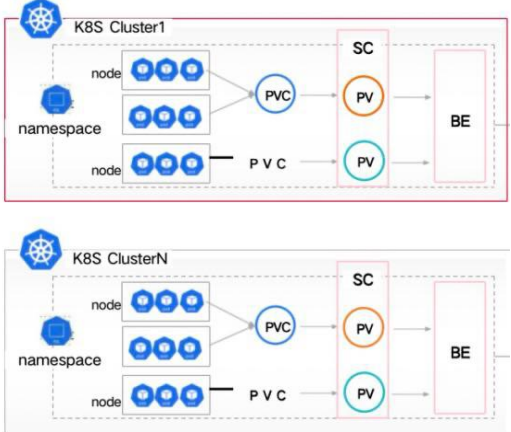

**Figure 3.** Cloud-native storage isolation

Microservices and containers in a cloud-native architecture can scale and scale independently to accommodate changes in traffic. In addition, the cloud native architecture also supports a variety of programming languages and frameworks, can flexibly choose their own technology stack, its high scalability and agility, shape the cloud native application cluster in the high availability, according to the characteristics of the business, if the business read and write request requirements are not high, can be adopted to achieve "by drift instead of cutting", POD drift instead of master/slave switching, single copy instead of multiple copies. And usability does not degrade; If services have high requirements on database read/write performance, connection number, and concurrency, you can use the primary/secondary cluster mode to implement read/write separation, personalized data initialization design, and strong data consistency.

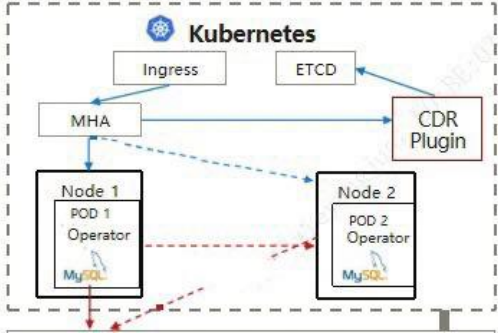

**Figure 4.** The cloud native database is highly available

The traditional MySQL high availability scheme completely relies on middleware for fault detection and master-slave switchover, and the pull-up time is strongly related to the data difference between the master and slave nodes. However, in the high availability scheme of cloud native database, due to the extreme drift capability of K8S, the data bank pull-up time is limited by the StatefulSetSet mode itself. When the faulty Node needs to be manually deleted, the middleware is combined with the storage CDR, and the storage initiates the rapid drift of the faulty [14]POD. Implements HA switchover in single-copy mode within 30s.

In conclusion, the methodology outlined here presents a robust framework for addressing the evolving landscape of financial information security within cloud-native environments. By leveraging a combination of cloud-native storage solutions and intelligent prediction technologies, financial institutions can effectively mitigate security risks while optimizing data management and operational efficiency.

Furthermore, the integration of multi-tenancy capabilities and tenant rights mechanisms ensures that cloud-native storage remains secure and accessible, even in shared environments. With the continuous evolution of cloud-native technologies and security practices, financial institutions are poised to strengthen their resilience against emerging threats while maximizing the benefits of cloud computing in the financial sector.

## 4. Conclusion

Cloud-native agile deployment and delivery, increased reliability and fault tolerance, increased scalability and flexibility, and increased security. This kind of future-proof architecture pattern can help enterprises build efficient, reliable, flexible and secure application ecosystems. From the perspective of application architecture, cloud native can easily support microservice architecture to realize application modernization and more flexible response to changes and elastic expansion. In terms of software lifecycle management, cloud-native technology can help implement DevOps and other best practices into applicable standardized tools and frameworks, greatly improve development efficiency, accelerate iteration, help developers and enterprises more easily move to the cloud and cloud, and enable applications to dynamically migrate in their own data centers and clouds.

With the continuous development of the financial market and the rapid popularization of cloud computing technology, the field of financial information processing is facing unprecedented challenges and opportunities. By combining cloud computing with financial markets, we can enable more efficient and accurate processing of financial information and effectively address security and privacy challenges. First, the application of intelligent forecasting and evaluation technology will greatly improve the ability of financial institutions to process and manage data, so as to better grasp business opportunities and risks. Secondly, through the adoption of the cloud computing model, financial institutions can quickly obtain and release computing resources, which greatly improves the flexibility and response speed of the business. At the same time, the resource integration, strong anti-attack capability and high scalability in the cloud computing environment also provide strong support for the security protection of financial information. However, with the growing demand for financial information processing, we are also faced with challenges such as data security and privacy protection. Therefore, financial institutions need to strengthen the control and protection of data security in the cloud computing environment, and take a series of effective security measures to ensure the security and stability of financial information. Through continuous technological innovation and cooperation, the financial industry will be able to better cope with future challenges, achieve the goal of digital transformation, and inject new vitality and momentum into the development of the financial market.